# Diffusion mediated photoconduction in multi-walled carbon nanotube films

**Biddut K. Sarker, M. Arif, Paul Stokes and Saiful I. Khondaker** [*]
Nanoscience Technology Center and Department of Physics, University of Central Florida, Orlando, Florida 32826, USA.

* To whom correspondence should be addressed. E-mail: saiful@mail.ucf.edu

**Abstract**
We investigated the mechanism for photoconduction in multi-walled carbon nanotube (MWNT) film of various electrode separations upon near infrared illumination. In addition to observing strong dependence of photocurrent on the position of the laser spot, we found that the time constant of the dynamic photoresponse is slow and increases with increasing electrode separations. The photoconduction mechanism can be explained by the Schottky barrier modulation at the metal-nanotube film interface and charge carrier diffusion through percolating MWNT networks.

(Some figures in this article are in colour only in the electronic version)

**1. Introduction:**

Carbon nanotubes (CNTs) are considered to be promising building blocks for nanoelectronic and optical devices due to their special geometry, high electrical conductivity, and exceptional mechanical and optical properties [1-2]. In particular, photoresponse studies of pure CNT films and CNT/Polymer composites have attracted tremendous attention because of their promising applications for bolometer, position sensor and photovoltaic devices [3-6]. Itkis *et al.* found that in single walled carbon nanotube (SWNT) film, the photoresponse was due to a bolometric effect, a change in conductivity due to heating of the SWNT network [3], while Levitsky *et al* showed that in SWNT film, molecular photodesorption to be responsible for change in conductivity upon near IR illumination [7]. Pradhan et al. [8] in SWNT/polymer composite also found that the photoresponse is bolometric. However, in these measurements, the size of the electrodes was either smaller than the laser spot size or the laser was positioned in the middle and authors did not check the effect of contacts.

Other studies in macroscopic SWNT films [9-14], multi-walled carbon nanotube (MWNT) film [4], and MWNT/polymer composites [15] with large electrode separation have shown that the photocurrent generation depends upon the position of the laser spot and maximum photoresponse occurs at the metal-CNT film interface. This effect has been explained using an exciton model where the absorption of light creates a bound electron hole pair and the Schottky barrier at the metal-CNT film interface helps to separate electrons and holes to induce a photo voltage [9]. Another common observation in all these studies was very slow time response (~1s) of photocurrent. The reason for the slow time response is a matter of considerable debate.

In this paper, we present a near-infrared photoresponse study of MWNT film with electrodes separation ranging from 2 mm to 50 mm to investigate the photoconduction mechanism in MWNT film. In agreement with previous reports, we also observed strong dependence of photocurrent on the position of laser spot with maximum photoresponse occurring at the metal MWNT interface. In addition, we found that the time constant of dynamic photoresponse at metal-film interface depends upon the electrode separation and that the time constant increases from 0.35 to 5.3 seconds as the electrode separation increases from 2 mm to 50 mm. While the photocurrent generation can be explained by Schottky barrier modulation at the metal-CNT film interface, the slow time response can be described by a model of the diffusion mediated conduction of charge carriers through many interconnected MWNTs.

## 2. Experimental details:

MWNT films were prepared using a drop cast technique. The MWNTs with a purity of >95% were purchased from Nanolab. The diameter and length of the as purchased MWNTs are 10-20 nm and 5-20 μm, respectively. MWNTs were dispersed into 1, 2 dichloroethane and sonicated for 3-4 hours in water bath kept at constant temperature of 10-15 $^0$C. The concentration of the solution was 1 mg/ml. After dispersion, appropriate amount of solution was drop cast onto a glass slide to make a thin layer of film. The slide was kept on a hot plate at around 40-50 $^0$C for 10-15 minutes to evaporate the solvent after which another layer of thin film was deposited. The resulting film had a thickness of ~ 40 μm. Finally, conducting silver paste was used to make pairs of electrodes of various separations $d$ = 2, 3, 5, 10, 20, 25, 40, 50 mm with a fixed width of 25 mm. Figure 1a show a field emission scanning electron microscope image of one of our film and figure 1b shows an optical micrograph of one of the samples with 10 mm and 50 mm electrode separations.

Figure 1c shows a schematic diagram of a final device and the electrical transport measurement setup. The room temperature dc charge transport measurements of the MWNT films was carried out in a probe station using a standard two-probe technique both in the dark and under illumination by a laser spot positioned at three different locations: L corresponds to illumination on the left electrode/film interface, M is between the electrodes in the middle of the sample, and R is the right electrode/film interface. The near IR photo source consists of a semiconductor laser diode with peak wavelength of 808 nm (1.54 eV) driven by a Keithley 2400. The spot size of the laser was approximately 10 mm long and 1 mm wide. The photo intensity was monitored with a calibrated silicon photodiode (Thorlabs S121 B). Unless mentioned otherwise, the power intensity of the laser was ~ 4 mW/mm$^2$ at the distance it was placed from the sample (~20 mm). Photocurrent was measured by applying 1mV bias between electrodes, measuring current by illuminating the film with NIR and subtracting the dark current.

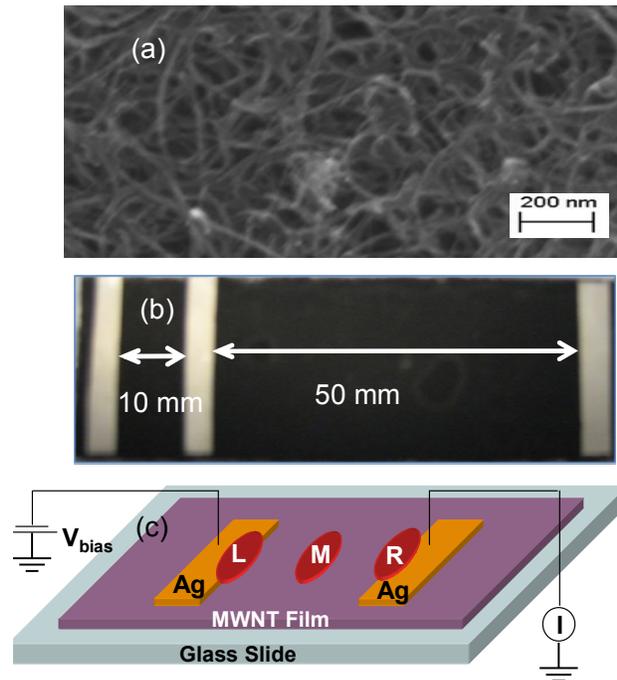

Figure 1. a) Scanning electron micrograph of a MWNTs film (b) An optical micrograph of one of the samples showing two pairs of electrodes. (c) Schematic diagram of the device and electric transport measurement set up. The spacing between the electrode varied from 2 – 50 mm, and IR laser wavelength is 808 nm. L, M and R marks the position of the laser.

Data was collected by means of LabView interfaced with the data acquisition card and current preamplifier (DL instruments: Model 1211) capable of measuring sub pA signal.

### 3. Results and Discussions:

Figure 2 shows a typical photoresponse curve for one of our MWNT films with electrode separation d = 10mm, where we plot photocurrent as a function of time (*t*) when the laser spot was positioned at L, M, and R and was turned on and off every 100 s interval. It can be seen that the photocurrent strongly depends on the position of the laser spot. When illuminated at position L there is an increase in photocurrent. When shined at position M, there was almost no photocurrent generation, whereas position R shows a decrease in photocurrent when illuminated by the NIR source. It can be seen that the on and off current is completely reproducible over several cycles.

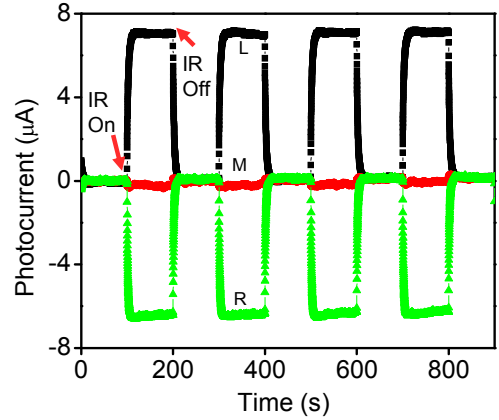

Figure 2. Representative photocurrent, as a function of time for a film with 10 mm electrode separation under IR illumination at positions L, M, and R ($V_{bias}$=1mV). The IR laser is turned on and off at every 100 s interval.

Similar position dependent behaviour of the photocurrent has been observed in all our samples with electrode separations ranging from 2 mm to 50 mm. The large enhancement of photocurrent at the metal – carbon nanotubes interface can be described by Schottky barrier model [9]. When the laser is shined at the left metal-nanotube interface, photons are absorbed by carbon nanotubes which in turn creates excitons (bound electron-hole pair). Some of these electrons have enough energy to overcome the barrier potential by tunnelling or thermal emission and fall into metal electrode leaving holes in the nanotube film. This induces a separation of electrons and holes at the interfaces and creates a local electric field. Therefore, a positive photocurrent generates at this interface. On the other hand, when the laser shines at right interface, the separation of electrons and holes also generate a local electric field, but in the opposite direction than that of left interface. Therefore the photocurrent is negative with almost same magnitude as of left electrode. However, when the laser light shines at the middle part of the sample M, electron-hole pairs are also generated but because of the absence of interface they do not get separated and no local electric field is created at this point. So, almost zero photocurrent is seen at the middle position.

Figure 3a shows a representative plot of photocurrent vs. time when illuminated at position L

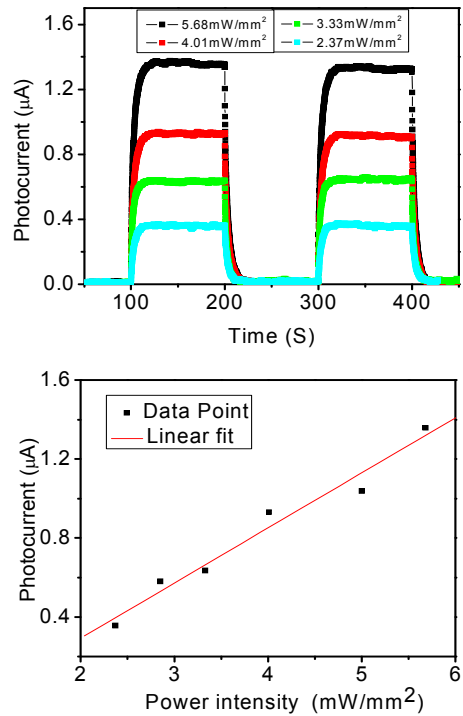

Figure 3. (a) Photocurrent versus time for a few different laser intensities. (b) Dependence of Photocurrent of the MWNT film on the laser intensity.

for another sample with electrode separation 25 mm for a few laser intensities (2.37, 3.33, 4.01, and 5.68 mW/mm² from bottom to top). The intensity of laser light was changed by changing the height between the sample and laser source. The plot is shown for two cycles of the laser being turned on and off at every 100 s intervals. In figure 3b we plot the photocurrent versus laser power intensity for the same sample shown in figure 3a for all the laser intensities. The solid line is a linear fit of the data which show that the photocurrent increases linearly with intensity. Similar observation was reported for SWNT films [7]. When the intensity of the laser light is higher, more photons are absorbed by the carbon nanotubes and generate more excitons. So a greater number of electrons have the probability to overcome the Schottky barrier, generating a larger photovoltage. On the other hand, when intensity of laser light is low, a smaller photovoltage is generated.

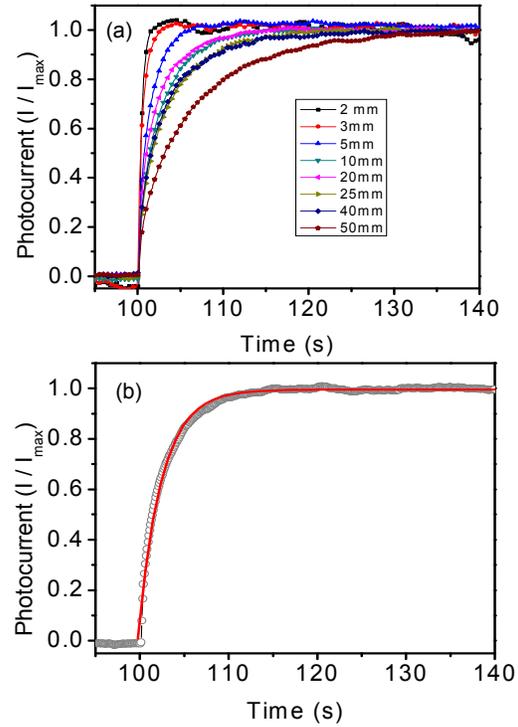

Figure 4. Time response of the photocurrent. (a) Raising part of normalized photocurrent as a function of time for all film with electrode separations of 2 – 50 mm. The laser light was positioned at left MWNTs/electrode interface. (b) Sample with electrode separation 10mm. The open circles are data and the solid line is the exponential fit with a time constant $\tau$ = 2.65 s.

We now investigate the time response of the photocurrent. Figure 4a shows a plot of the rising part (first 40 s) of the normalized photo-current ($I/I_{max}$) versus time for all of our MWNT films with electrode separations of 2, 3, 5, 10, 20, 25, 40 and 50mm when illuminated at left electrode. The top curve is for d = 2 mm and the bottom is for d = 50 mm. Two features can be noticed from this data: (i) the response time, time taken to reach maximum photocurrent is rather slow and (ii) the response time increases with increasing separation. The dynamic response to the NIR source can be well described by $I = I_0[1 - \exp(-(t-t_0)/\tau)]$, where $\tau$ is time constant, $t_0$ is the time when NIR is switched on, and $I_0$ is the steady state photo current. Figure 4b shows a fit of this equation for one of the sample with electrode separation of 10 mm. Open circles are the experimental points and the solid line is a fit to the above equation. From this fit we obtain $\tau$ =2.65 s. Similar fits were done for all the samples and the measured time constants were 0.35, 0.62, 1.50, 2.65, 3.39, 4.13 and 5.27 seconds for 2, 3, 5, 10, 20, 25, 40 and 50 mm electrode separations respectively. From here, we conclude that the time constant increases with increasing electrode separations. Similar increases in time constants were also obtained for the decaying part when the laser was switched off and for the right electrode-CNT film interface.

There is a lot of debate about the origin of slow time response of photocurrent in CNT films. Previous studies in SWNT films have shown that bolometric effect [3] and molecular photodesorption [7] can explain the photoresponse and slow time response in photocurrent. We rule out both these effects in our film because the positive and negative response at two different interfaces cannot be explained by these models. Recently, it has been shown that in SWNT film the slow time response can be explained by carrier diffusion model [12]. According to the diffusion model, considering a parabolic impurity density distribution, the time constant can be described as [12, 16]

$$\tau = \frac{d^2}{2D_p(r+1)}\left[1 - \frac{I_{(1/2)(r+3)}(d/L_p)}{I_{(1/2)(r-1)}(d/L_p)}\right] \quad (1)$$

where $d$ is the electrode separation, $L_p$ is diffusion length, $D_p$ is diffusion coefficient of hole, $I_n$ is modified Bessel of the first kind of order $n$ and $r \sim d^\alpha$ is a real constant for a parabolic impurity density distribution function. According to this model, the time constant should increase with increasing electrode separations. In order to see whether the diffusion model can describe the slow time response in our MWNT networks, we plotted the time constant as a function of electrodes separation in figure 5 and fitted the data with the above equation. The black squares are the measured time constants while the solid curve is a fit to the diffusion equation for charge carrier using $L_p$=1 mm, $D_p$=0.01cm$^2$/Vs and α=1.4 as fitting parameters. These parameters are similar to what was obtained for SWNT networks. It can be seen from figure 5 that the experimental data can be fitted reasonably well with the diffusion model. Therefore, we conclude that the slow time response in our film is due to the diffusion of free charge carriers that was created at the metal-MWNT film interface. In other words, the slow response is due to the diffusion mediated charge transport through many interconnected individual MWNTs.

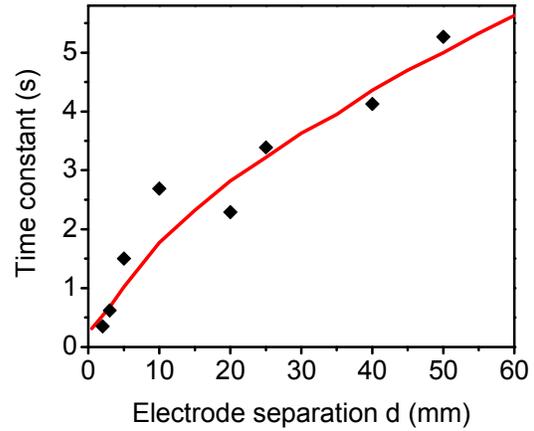

Figure 5: Variation of time constant with electrode separation. The black squares are the measured time constant and the solid line is a fit to the diffusion model.

### 4. Conclusions:

In conclusion, we presented NIR photoresponse study of MWNT films with various electrode separations. We found that the photoresponse is position dependent with highest photoresponse occurring at MWNT/metal interface and is consistent with the model of Schottky barrier modulation for photocurrent generation. The time constant for dynamic photoresponse increases with increasing electrode separations and can be explained by the diffusion of charge carriers through percolating MWNT interconnects. Our results presented here may have important implications on the use of CNT thin films for photodetectors and photovoltaic devices.

### Acknowledgement:

This work is supported in part by US National Science Foundation under grant ECCS 0801924.